\documentstyle[12pt,epsf] {article}

\newcommand{\ai}{{\it ab initio }}
\newcommand{\ea}{{\it et al.}}

\begin{document}

\begin {center}
{\bf {Thermodynamics of Na$_8$ and Na$_{20}$ clusters %
studied with {\it ab initio} electronic structure methods}}

Abhijat Vichare and D. G. Kanhere\footnote{(amv, kanhere)@unipune.ernet.in}

Department of Physics, University of Pune, Ganeshkhind, Pune 411
007, INDIA

S. A. Blundell\footnote{sblundell@cea.fr}

D\'{e}partement de Recherche Fondamentale sur
la Mati\`{e}re Condens\'{e}e, CEA Grenoble \\
17, rue des Martyrs, F-38054 Grenoble CEDEX 9, France

\end {center}
\begin{abstract}

    We study the thermodynamics of Na$_8$  and Na$_{20}$ clusters using
    multiple-histogram methods and an {\em ab initio} treatment of the
    valence electrons within density functional theory.  We consider the
    influence of various electron kinetic-energy functionals and
    pseudopotentials on the canonical ionic specific heats.
    The results for all models we consider show qualitative
    similarities, but also significant temperature shifts from model
    to model of peaks and other features in the specific-heat curves.
    The use of   phenomenological
    pseudopotentials  shifts the melting peak  substantially ($\sim$
    50--100~K)  when compared to \ai results.   It is argued that the
    choice  of a good  pseudopotential and  use of better electronic
    kinetic-energy functionals has the potential for performing large
    time scale and large sized thermodynamical simulations on clusters.

\end{abstract}          
\section{Introduction}

\label{introduction}

    The physics of finite-sized systems such  as clusters continues to
    invoke  considerable  interest both in    theory and experiment.  A
    particularly  intriguing  and poorly  understood  phenomenon is the
    melting behavior of such  finite-sized systems.  Only recently have
    Haberland   \ea \cite {schmidt}  succeeded   in measuring the  heat
    capacity of free (i.e. unsupported) Na$_{n}^{+}$ clusters, with $n$
    ranging    from  70 to  200   atoms.    Interestingly, they  find a
    nonmonotonic behavior of the  melting temperature as a function  of
    cluster size,   with pronounced maxima  at $n=57$  and  142.  These
    sizes correspond neither to  closed-shell Mackay icosahedra ($n=55$
    and 147) nor   to closed shells of  valence   electrons ($n=59$ and
    139), but are intermediate between the two.  This clearly indicates
    that both geometric and  electronic shell effects contribute to the
    melting phenomenon in a rather subtle manner.

    Prior to  this  measurement, there have  been  a  few experimental
    studies on melting  of clusters. Martin \ea \cite{martin} reported
    measurements on  the melting temperature   of Na clusters for  the
    sizes of  the  order  of   thousands of  atoms and   their results
    indicated that the  melting temperatures increased with size,  but
    had  not    reached the  experimental   bulk  value.    Peters \ea
    \cite{peters} have  noted  the existence   of  surface melting  on
    supported Pb nanoparticles using X-Ray diffraction.

    Clearly, the melting behavior   at small sizes is  cluster specific
    and   dependent  on the  nature  of   the  electronic structure and
    geometry.  Further,   the   transition found   by   Haberland   \ea
    \cite{schmidt} is   not sharp  and has   a broadened   peak  in the
    specific heat  with a  width of  approximately  40 K. The  expected
    monotonic increase of melting   point has been  seen only  for very
    large clusters containing upwards of several thousand atoms. On the
    theoretical    side, much   insight   into   the finite-temperature
    properties  has been  gained   via  molecular  dynamics  (MD)   and
    Monte-Carlo (MC) numerical simulations.  Most of these  simulations
    have been carried out  using classical empirical two-body potential
    functions, \cite{berry}  mostly   of  Lennard-Jones  (LJ)  or Morse
    type. These studies  have revealed that   small clusters exhibit  a
    melting transition   over  a broad temperature range,   unlike bulk
    systems, and have broad   heat-capacity curves. In  addition,  they
    also  exhibit a variety of other  phenomena  such as  isomerization
    (including surface isomerization) and  surface melting, which   are
    generically referred to  as ``premelting'' phenomena. Some clusters
    also  exhibit  coexistence  of  liquid-like and   solid-like phases
    within the melting temperature range.

    MD  and MC  simulations  have also been  reported using  classical
    embedded-atom  potentials, such as the single-moment approximation
    (SMA),   \cite{bulgac,li}  which contain    approximations  to the
    $N$-body forces found in metallic systems like Na clusters.  Calvo
    and      Spiegelmann \cite{calvo,calvo2}  performed      extensive
    simulations on Na clusters with from 8 to  147 atoms using the SMA
    potential of Li  \ea\, \cite{li} with a view  to  probe the melting
    phenomena    of small  Na   clusters.  They  find  that premelting
    phenomena dominate the melting process  at small cluster sizes ($n
    < 75$),  while  the  larger   sizes exhibit  a  preference for   a
    single-process melting. They also  observe that the nature  of the
    ground   state  is  critical   to    the  thermodynamics of    the
    cluster. However, as they  clearly point out, their simulations do
    not incorporate the electronic structure effects directly.

    An alternative approach for metallic clusters is that of Poteau \ea\,
    \cite{poteau,poteau2}  who developed a tight-binding Hamiltonian to
    incorporate    quantal   effects    approximately.   They    use  a
    H\"{u}ckel-type Hamiltonian  and MC to  sample the  phase space for
    small Na clusters  with 4, 8, and  20 atoms. Calvo and  Spiegelmann
    \cite{calvo2} have performed  more extensive calculations for sizes
    up to 147 atoms with the  same potential.  However, during the last
    decade   developments in  \ai   methods have  opened  up  practical
    possibilities  of  performing  accurate  simulations  by  combining
    density functional theory (DFT) with classical  MD or MC.  The most
    accurate form of DFT is the Kohn-Sham (KS) formulation.    Although
    these  methods  have  been  used  to  investigate   the  structural
    properties with remarkable  success, relatively few applications of
    such \ai  methods have been to the  simulation of melting. Jellinek
    \ea     \cite{jellinek}   have,   however,     combined   a  hybrid
    Hartree-Fock/DFT     method  with   MC  sampling      to study  the
    thermodynamics of Li$_{8}$.

    Although  it is most  desirable to have  a full quantum mechanical
    treatment of electrons, as in the KS method, such simulations turn
    out  to be   expensive.  It is   also  to  be  noted  that typical
    simulation times used in purely  empirical potential MD are of the
    order of a few 100 ps or  more per energy point.  Considering that
    the most relevant sizes for  the experiment  are  in excess of  50
    atoms, the full \ai simulation may  turn out to be practically too
    expensive.\cite{ks_order_of_magnitude}  Hence   approximate
    methods leading  to    practical and fast  algorithms   have  been
    developed.  One such   technique is  density-based (DB)  molecular
    dynamics, where the electronic kinetic energy is approximated as a
    functional  of density  only.   For example,  Vichare and  Kanhere
    \cite{our_al13} performed  \ai simulations on an Al$_{13}$ cluster
    to investigate its  melting behavior. The DB  method has also been
    used by Aguado \ea \cite{1:aguado,2:aguado}  to study the  melting
    of Na   clusters ranging  in size from   8  to 142   atoms.  Their
    simulations are  of constant  total  energy type  using  empirical
    pseudopotentials with  up to 50  ps of observation time per energy
    point  for small clusters (8 and  20 atoms), and up   to 18 ps per
    energy point for larger ones.  Another approach  is that of Blaise
    \ea\,\cite{steve} who carried  out DB simulations for Na clusters
    up to size 274,  but using soft, phenomenological pseudopotentials
    rather  than \ai  pseudopotentials.    In addition  to  permitting
    significantly     longer  observation      times,    these    soft
    pseudopotentials were shown  to reproduce well properties  such as
    the  volume and surface  energies,   ionization energies, and  the
    frequency   of    collective   ionic   monopole   and   quadrupole
    oscillations.

    These above-mentioned studies on Na clusters bring out a number of
    issues which need further investigation.  Clearly, it is desirable
    to have  both long simulation times  and a full quantum mechanical
    treatment  of electrons.   Full  KS being  expensive, however, the
    attractive propositions   of  DB    or  soft pseudopotentials   as
    practical alternatives for the simulation  of such systems need to
    be assessed as to  their quality.  This is particularly  important
    because  of  some discrepancies seen   in the  above studies.  For
    example, in     the case  of  Na$_{8}$,   Calvo   and  Spiegelmann,
    \cite{calvo,calvo2}  using   an SMA   potential,   find that  the
    canonical specific heat  is broad in  nature and shows a flattened
    peak between  about 110 K  to 250  K.  For  the same cluster,  the
    tight-binding potential \cite{calvo2,poteau} leads to a less broad
    peak,  with a width of  about 70 K and  peaking at 160 K. However,
    the  microcanonical   specific   heat  obtained  by   Aguado   \ea
    \cite{1:aguado} for Na$_{8}$ in a constant-energy DB study shows a
    peak at a much lower value of  110 K and  is sharp with a width of
    less than 30~K.  This   is in  qualitative disagreement with    
    the SMA and tight-binding results.

    Further, there  is  a difference  in  the  way  the data  has  been
    analyzed by these workers. Aguado \ea \cite{1:aguado,2:aguado} have
    used the  traditional   trajectory-based analysis,  which  uses the
    caloric   curve supplemented   by    Lindemann  type  criteria  for
    identifying the transition.  Since  the transition is never  sharp,
    such an analysis may not  lead to  an unambiguous determination  of
    the melting  temperature. In addition,    the observation times  of
    Aguado \ea are  significantly  less than  those  used by Calvo  and
    Spiegelmann \cite{calvo,calvo2} in their SMA or tight-binding work.
    In fact,  it  is desirable  to calculate appropriate  thermodynamic
    indicators such   as the ionic entropy   and the specific  heat. In
    Refs.~\cite  {calvo,calvo2,poteau},   the authors   have  used  the
    multiple histogram  (MH)  technique \cite{ferrenberg,labastie}   to
    extract the entropy and the specific heat from the simulation data,
    as we do here.

    In the present work, we  therefore examine the melting of Na$_{8}$
    and Na$_{20}$  clusters with a view  to resolving these issues. We
    have carried out the following simulations on  Na$_{8}$: a full KS
    (orbital-based)   simulation  using  \ai  pseudopotentials;  a  DB
    simulation, where the  electronic kinetic energy  is approximated,
    but with identical  pseudopotential  and time scales; and  both KS
    and  DB  simulations    with  soft pseudopotentials.     The  same
    simulations  have  been carried    out  for Na$_{20}$,   with  the
    exception of the full KS simulation with \ai pseudopotentials.  In
    all the cases  we have  calculated the  entropy and  the canonical
    specific  heat  via the  MH  method, as  well  as  the traditional
    indicators like the RMS bond  length fluctuation and mean  squared
    displacements.

    In  the next section, we  briefly  describe the formalism, analysis
    methods, and  numerical   details.  In  Section   \ref{results}, we
    present  our   results and discuss  them  in  the light  of earlier
    studies. Finally,  our    conclusions  are presented    in  Section
    \ref{conclusions}.

%
%
\section{Method}

\label{method}

   Following the usual procedure in  DFT,\cite {payne}  we write the total
   energy of a system of $N_{a}$ stationary Na$^{+}$ ions with  coordinates
   $R\equiv \{{\bf R}_{i}\}$ and $N_{e}$ valence electrons  as a functional
   of the electron density $\rho \equiv \rho ({\bf r})$
   \begin{equation}
   E_{{\rm pot}}[\rho ,R]=T[\rho ]+E_{{\rm ext}}[\rho ,R]+E_{H}[\rho
   ]+E_{xc}[\rho ]+E_{ii}[R]\,,  \label{eqn:epot}
   \end{equation}
   where $T[\rho ]$ and $E_{H}[\rho ]$  are the kinetic and Hartree energy,
   respectively, of the valence electrons, $E_{{\rm  ext}}[\rho ,R]$ is the
   interaction energy  of the valence  electrons  with the ions,  evaluated
   using   the pseudopotential formalism, $E_{xc}[\rho  ]$  is the electron
   exchange-correlation energy  in the local  density approximation  (using
   the parametrization of Perdew and Zunger\cite{perdew}), and $E_{ii}[R]$
   is the ion-ion interaction energy. In the  standard KS approach, $T[\rho
   ]$ is expressed as a sum of expectation values, over each KS orbital, of
   the electron  kinetic-energy operator $-(1/2)\nabla ^{2}$.  In contrast,
   in the  DB approach $T[\rho ]$ is  expressed as a  functional of $\rho $
   only,  without introducing orbitals,   leading  to  a faster though   in
   practice  less accurate calculational  scheme.  For each approach we use
   either {\em ab  initio} (AI) pseudopotentials or  soft, phenomenological
   (SP) pseudpotentials.  We consider  two forms for  $T[\rho ]$ in the  DB
   approach: in our  DB-AI approach we use  a functional form proposed  for
   clusters,\cite{ghosh,dbmd} while in  our DB-SP approach we take $T[\rho
   ]$ as a sum of the Thomas-Fermi energy and  a scaled Weizs\"{a}cker term.
   \cite{steve}
 
   The {\em ab initio} pseudopotentials used in the KS-AI and
   DB-AI  approaches  are  those
   proposed  by Bachelet,  Hamann  and Schl\"{u}ter.\cite{bhs}  The  soft,
   phenomenological pseudopotential used in  the KS-SP and DB-SP approaches
   is given by \cite{steve}
   \begin{equation}
   V_{{\rm soft}}(r)=\left\{
   \begin{array}{ll}
   -\frac{1}{r}\,, & r>r_{c} \\
   -\frac{1}{6r_{c}}\left[ 7-\left( \frac{r}{r_{c}}\right) ^{6}\right]
   \,, & r \leq r_{c}\,,
   \end{array}
   \right.   \label{eqn:vflat}
   \end{equation}
   for a single Na$^{+}$ ion at  the origin, where $r_{c}=3.55a_{0}$ in the
   DB-SP approach and $r_{c}=3.7a_{0}$ in the KS-SP approach. The choice of
   $r_{c}$ for the DB-SP approach follows from  a fit to volume and surface
   energies,\cite{steve} while   for   the KS-SP approach  the   choice of
   $r_{c}$  ensures  close   agreement   with   ionization   energies   and
   dissociation  energies given by an  {\em ab initio} pseudopotential, for
   small clusters in the size range $n=3$ to 8. Use of the phenomenological
   pseudopotential  permits  a  larger grid  step  size or  equivalently  a
   smaller plane-wave energy cut-off, thus leading  to a faster solution in
   either the KS or DB formalisms.  
   The Car-Parinello (CP) algorithm \cite{cp} was used in the DB-AI and DB-SP
   schemes, while the damping scheme proposed by Joanopoulous {\it et al.}
   \cite {joan} was used to minimise the electronic degrees of freedom in 
   KS-AI.

   The trajectories collected were analyzed using traditional indicators of
   melting like the rms bond-length fluctuation, defined as
   \begin{equation}
   \delta _{{\rm rms}}=\frac{2}{N_{a}(N_{a}-1)}\sum_{i<j}\frac{\left(
   \langle r_{ij}^{2}\rangle _{t}-\langle r_{ij}\rangle _{t}^{2}\right)
   ^{\frac{1}{2}}} {\langle r_{ij}\rangle _{t}}\,,  
   \label{eqn:rmsblf}
   \end{equation}
   where  $r_{ij}$  is the  distance   between   ions $i$   and $j$,    and
   $\left\langle     \ldots \right\rangle     _{t}$     denotes    a   time
   average. According to the Lindemann criterion, a system may no longer be
   considered to be solid if $\delta  _{{\rm rms}}$ is  greater than 0.1 to
   0.15. Short  time  averages over the trajectory  data,  e.g.\ over  data
   points corresponding to 1 ps, 2 ps, 5 ps, etc., were evaluated to obtain
   the dependence  of $\delta _{{\rm  rms}}$  on the duration  of the  time
   average.  Another   indicator we have  used  is  the  mean  square ionic
   displacement, defined as
   \begin{equation} 
   \langle {\bf r}^{2}(t)\rangle =\frac{1}{Nn_{t}}\sum_{m=1}^{n_{t}}
   \sum_{i=1}^{N_{a}}\left[ {\bf r}_{i}(t_{0m}+t)-{\bf
   r}_{i}(t_{0m})\right] ^{2}\,.  
   \label{eqn:msq}
   \end{equation}
   We have set the total number of time-steps $n_{t}$ used in the time
   average  to $n_{t}=n_{T}/2$, where  $n_{T}$ is the total simulation
   time (usually about 50 ps).

   A more  complete thermodynamic analysis  of  the simulations is possible
   using the multiple histogram method (MH),\cite{ferrenberg,labastie} and
   all  simulations were   analysed  using  this  method. It  requires  the
   configurational   energy,   which  corresponds   here  to the  classical
   potential energy $E_{{\rm pot}}[\rho ,R]$ of Eq.\ (\ref{eqn:epot}), over
   various points in the ionic phase space accessed by the system along the
   trajectory. This  is used to evaluate  the classical ionic  density   of
   states ${\Omega }(E)$, and thereby the ionic   entropy $S(E)=\ln {\Omega
   }(E)$, as  well as  the partition  function  via a least-squares fitting
   procedure.  The sampled values of  the configurational energy are fitted
   to the theoretical probability distribution  and the fitted coefficients
   are  then  used  to evaluate the   various  thermodynamic functions.  We
   consider in particular the canonical specific heat, defined as usual by
   \begin{equation}
    C=\frac{\partial {U}}{\partial {T}}\,,  \label{eqn:specheat}
   \end{equation}
   where $U=\langle E_{{\rm  pot}}+E_{{\rm  kin}}\rangle _{T}$  is the
   average    total  internal  energy  in  a    canonical ensemble  at
   temperature $T$. We here exclude the  contribution of the center-of-mass 
   motion to the ion kinetic energy $E_{{\rm kin}}$, so that from
   the equipartition theorem
   \begin{equation}
   \langle E_{{\rm kin}}\rangle _{T}=\frac{3}{2}(N_{a}-1)k_{B}T\,.
   \label{eqn:ekin}
   \end{equation}
   The canonical probability distribution for observing a total energy
   $E$ at temperature $T$ is given by the usual Gibbs distribution
   \begin{equation}
   p(E,T)=\frac{1}{Z(T)}\Omega (E)\exp \left( -\frac{E}{k_{B}T}\right)
   \,, 
   \label{eqn:gibbs}
   \end{equation}
   with  $\Omega (E)$ the classical  density  of states extracted from
   the  MH fit,    and  $Z(T)$ the    normalizing  canonical partition
   function.  Note that although here we  shall discuss results in the
   canonical   ensemble, once $\Omega   (E)$  is known,  one may  also
   evaluate properties in   the microcanonical ensemble,  such as  the
   microcanonical temperature $T(E)$
   \begin{equation}
   \frac{1}{T(E)}=\frac{\partial }{\partial {E}}\ln \Omega (E)\,.
   \label{eqn:mh_temp}
   \end{equation}


   Simulated annealing   was used  to  obtain the   ground-state ionic
   structures from a randomly  chosen  initial configuration for  each
   cluster.  For Na$_{8}$ the ground-state geometry is found to have a
   dodecahedral $D_{2d}$ symmetry in both the KS and DB formalisms and
   for both  the AI\ and  SP  pseudopotentials, in agreement with  the
   structure found  by R\"{o}thlisberger and Andreoni \cite {andreoni}
   in a  KS approach. 
   For  Na$_{20}$ in  the DB  formalism,  we find a ground state 
   consisting  of a double  icosahedron with a single cap on its waist, 
   which is   the second of  the    two  structures  found in 
   Ref.\ \cite{andreoni}. In the KS formalism,    the  ground  state for
   Na$_{20}$  is  a  double icosahedron missing  one end cap and  with 
   two caps on the waist, in agreement with Ref.\ \cite{andreoni}.   
   Our DB structures agree with those found by Aguado
   {\em et. al}.\cite{1:aguado}

   We have  considered two approaches  to the statistical     sampling
   of the ionic phase space, required as input  to the MH analysis. In
   each approach the clusters  are effectively heated slowly  from the
   ground-state structure at 0~K to a  liquid-like state at upwards of
   250~K. The  first approach involves  a canonical sampling    of the
   phase space and was used with the AI  pseudopotentials  in both the
   KS and  DB  schemes.  Successive  simulation temperatures  of 60~K,
   80~K, 100~K,   125~K, 150~K, 175~K,  200~K,  225~K,  and 250~K were
   chosen.  Each temperature was  maintained  within $\pm 10$~K  using
   velocity scaling,\cite{velref}   except  for the   60 K  and  80 K
   temperatures, where  the temperatures  were maintained  within $\pm
   5$~K.   The  total observation time  for both  KS   and DB is about
   57.5~ps  per temperature  point.  The   initial  condition at  each
   temperature  was taken     as  the final   state of   the  previous
   temperature, and the initial  1.25~ps of simulation time  were used
   to raise  the   previous  temperature. The   next 5~ps  were   then
   discarded to  allow for thermalization   of the  system at the  new
   temperature.  The analysis was  performed on the data corresponding
   to the last $\sim 50$~ps. The simulations for the clusters Na$_{8}$
   and Na$_{20}$ were performed  within  a cubical supercell  of  edge
   40~a.u.\ \cite{units} or  more. All Fourier space  evaluations were
   carried out   on a mesh   of  64$\times $64$\times $64  for  DB and
   48$\times $48$\times $48 for KS  with a cutoff  of about 21 Ry. The
   configuration energy  range was divided into  bins  whose width was
   chosen to give at least about 30  points for the lowest temperature
   distribution.   About 500 bins were    typically used to cover  the
   entire configuration  energy range.   The canonical specific  heats
   obtained using  the MH analysis were  then plotted as a multiple of
   their  value   $C_{0}$ at 0~K   given by $C_{0}=(3N_{a}-9/2)k_{B}$,
   which  is the  zero-temperature classical  limit  of the rotational
   plus vibrational specific heats.

   Our second approach consists of a microcanonical sampling    of the
   phase space, and  was used with the SP  pseudopotential in both the
   KS and DB schemes. Constant total energy simulations were performed
   at closely spaced values of the total energy, such  as to give good
   overlap of successive histograms  of the potential energy  $E_{{\rm
   pot}}$. The simulations were performed in order of increasing total
   energy,  with the  initial  condition  at  one energy obtained   by
   scaling the  velocities of the  final state of the previous energy,
   and 20--30  energy  points were used to   scan the  required energy
   range.    Each   energy point   consisted    of  from 50--100~ps of
   observation     time, of   which  5--10~ps    were  discarded   for
   equilibration. Several  scans of the entire  energy range were made
   in this way, giving {\em total} simulation times of about 15~ns for
   (Na$_{8}$, DB-SP), 5~ns for (Na$_{8}$, KS-SP), 6~ns for (Na$_{20}$,
   DB-SP), and   3~ns   for  (Na$_{20}$,  KS-SP).  The  microcanonical
   sampling   requires  a modified MH  analysis.\cite{labastie2} Note
   that DB-SP results have been  reported elsewhere,\cite{steve2} and
   are reproduced here for purposes of comparison.

   The dominant error in our specific-heat  curves is statistical, due
   to  the finite duration of the  sampling    of the  phase space. By
   adding extra data points, or complete additional scans of the whole
   temperature range,  to the MH\  analysis, we find the specific-heat
   curves to be stable to  about 10\% or  better, and the positions of
   peaks to be stable to  about $\pm 20$~K  or better. We take this as
   an   informal estimate  of  the   statistical  error.  However,  in
   dynamical simulations such as these, it may  be that some processes
   of importance (e.g.\ isomerizations) occur on a physical time scale
   rather longer  than we have  considered, so that we  have imperfect
   ergodicity; all we can   say is that  our curves  do appear   to be
   rather stable  on the time  scales that we  have considered. We are
   currently considering  recent Monte-Carlo sampling methods  such as
   the parallel tempering method,\cite{partemp} which are designed to
   overcome the problem of long time scales and improve ergodicity.


\section{Results}

\label{results}
\begin{figure}
    \epsfxsize=4in
    \epsfysize=3in
    \centerline{\epsffile{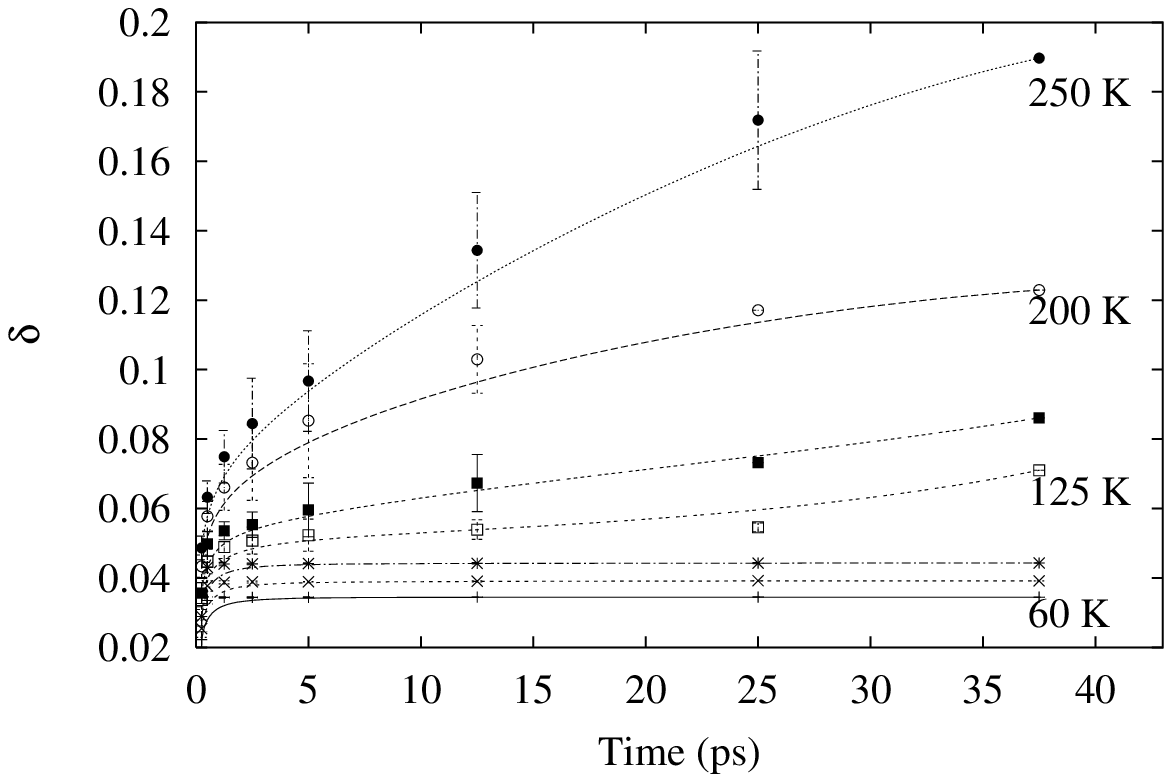}}   
    \caption  {The rms bond-length fluctuation of Na$_8$  simulated using
    the KS-AI model as  a function  of time  for various temperatures. 
    Note that the tendency to converge is faster at low temperatures.}
    \label {ks:na8:rms:time}
\end{figure}

\begin{figure}
    \epsfxsize=4in
    \epsfysize=3in
    \centerline{\epsffile{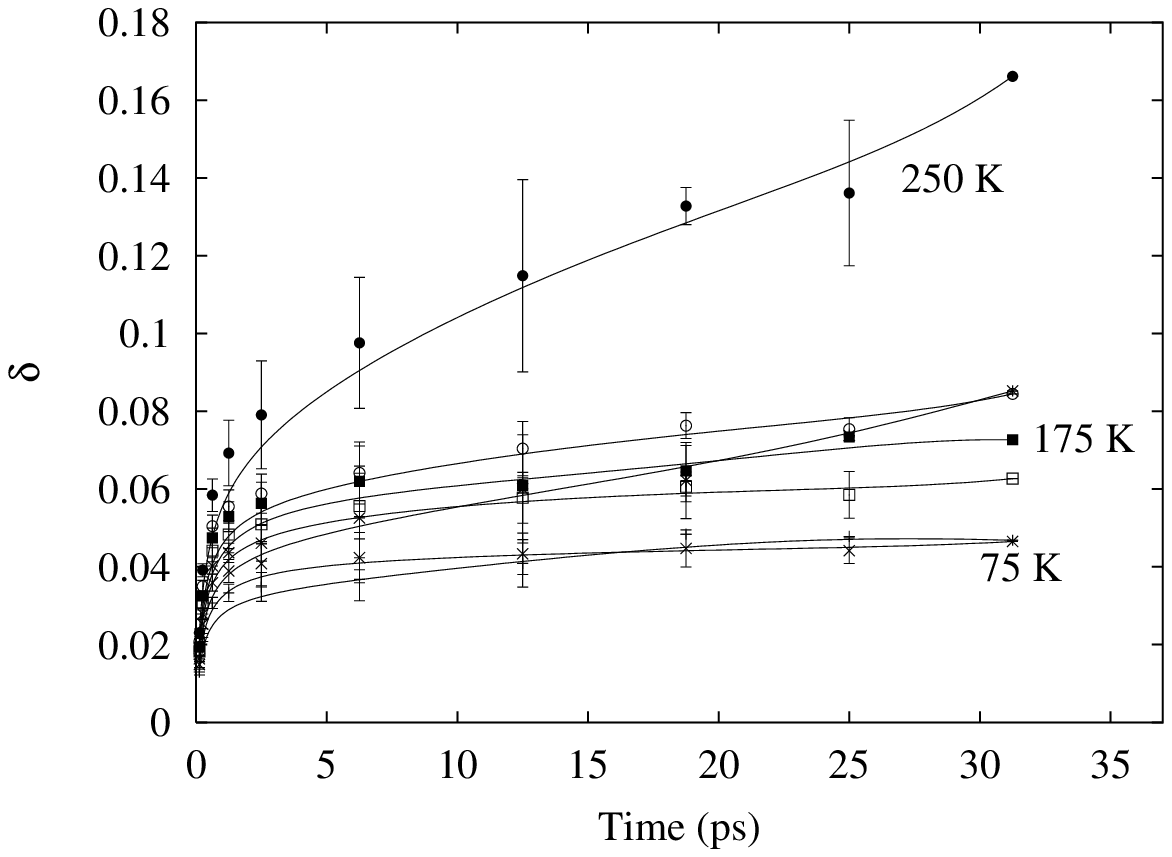}}   
    \caption  {The rms bond-length fluctuation of Na$_{20}$  simulated   using
    the DB-AI model as  a function  of time  for various temperatures. 
Note that the
    tendency to converge is faster at low temperatures.}
    \label {dbmd:na20:rms:time}
\end{figure}

\begin{figure}
    \epsfxsize=4in
    \epsfysize=3in
    \centerline{\epsffile{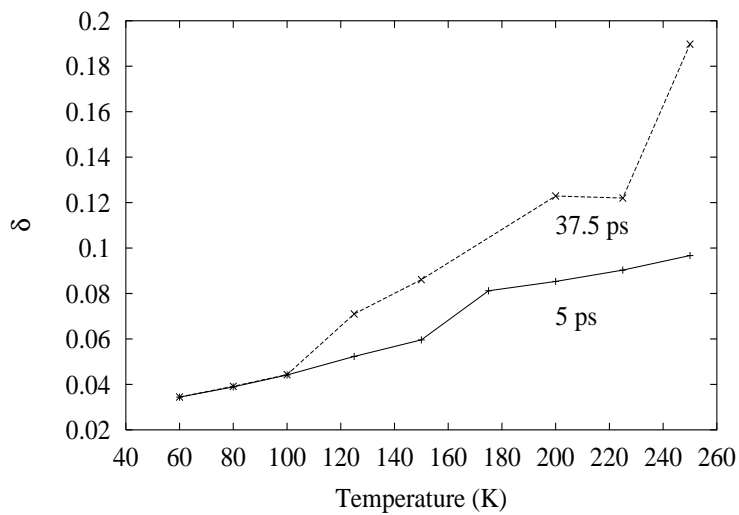}}   
    \caption  {The rms bond-length fluctuation of Na$_{\rm 8}$ simulated using
    the KS-AI model as a  function of temperature  over 5 ps and 37.5 ps.}
    \label {ks:na8:rms:temperature}
\end{figure}

We begin the discussion by considering some of the
conventional trajectory-based indicators of isomerization and melting. In
Fig.\ \ref{ks:na8:rms:time}, we show the rms bond-length fluctuations $%
\delta _{{\rm rms}}$ of Na$_{8}$ in the KS-AI model as a function of
simulation time, for different temperatures in the range 60~K to 250~K. The
figure makes it clear that for temperatures up to about $T=200$~K, 25~ps
are sufficient to converge the value of $\delta _{{\rm rms}}$, while for
higher temperatures of the order of 250~K or more, even longer simulation
times may be required. Similar behaviour is seen in Fig.\ \ref
{dbmd:na20:rms:time} for the 20-atom cluster simulated within the DB-AI
model. In Fig.\ \ref{ks:na8:rms:temperature}, we show $\delta _{{\rm rms}}$
averaged over 37.5~ps and over 5~ps, as a function of temperature. Note that
the 5~ps curve never crosses the Lindemann criterion of 0.1, while the
37.5~ps curve crosses the Lindemann criterion of 0.1 around 190~K. The behavior
of $\delta _{{\rm rms}}$ in DB-AI over identical simulation times is very
similar.

\begin{figure}
    \epsfxsize=4in
    \epsfysize=3in
    \centerline{\epsffile{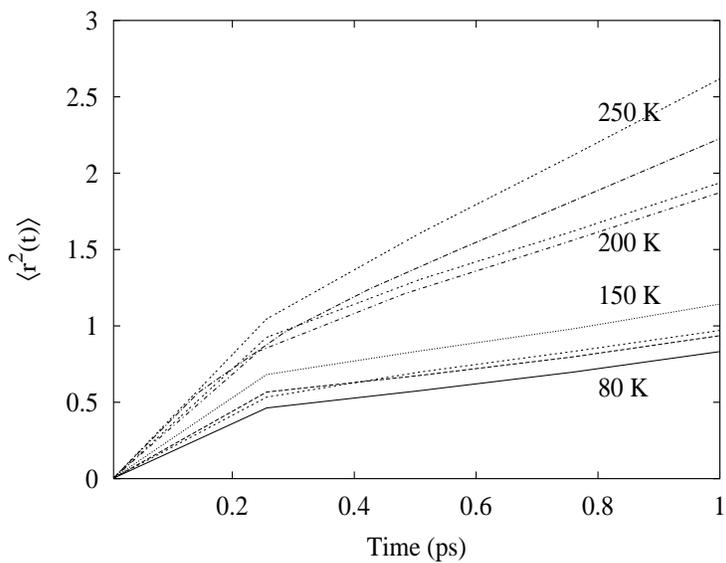}}   
    \caption {Mean square ionic displacements of Na$_{\rm 8}$ simulated using
    the KS-AI model at 1 ps time scale.}
    \label {ks:na8:1ps:msq}
\end{figure}

\begin{figure}
    \epsfxsize=4in
    \epsfysize=3in
    \centerline{\epsffile{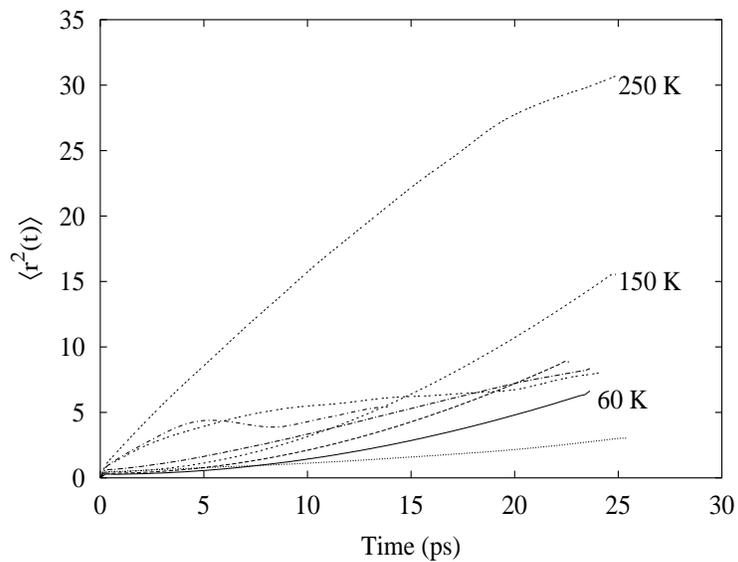}}   
    \caption {Mean square ionic displacements of Na$_{\rm 8}$ simulated using
    Kohn-Sham at 25 ps time scale.}
    \label {ks:na8:50ps:msq}
\end{figure}

The mean square ionic displacement $\langle {\bf r}^{2}(t)\rangle $ (\ref
{eqn:msq}) has also often been used as an indicator of isomerization or of a
solid-like to liquid-like transition. In Figs.\ \ref{ks:na8:1ps:msq} and \ref
{ks:na8:50ps:msq} we show $\langle {\bf r}^{2}(t)\rangle $ on different time
scales of 1~ps and 25~ps, respectively, for Na$_{8}$ in the KS-AI\ model.
One observes that $\langle {\bf r}^{2}(t)\rangle $ at low temperatures $%
T<100 $~K reaches a horizontal plateau for t$>\sim$ 0.25~ps, indicative
of a solid-like behavior in which atoms vibrate around fixed points with an
amplitude squared that increases in rough proportion to the temperature. On
the other hand, the rising curve for $T\geq 250$~K suggests a liquid-like
behavior with diffusion throughout the entire volume of the cluster. The
curve for $T\geq 250$~K would eventually reach a plateau with a $\langle
{\bf r}^{2}(t)\rangle $ value characteristic of the square of the linear
dimension of the cluster, but even at $t=25$~ps this plateau has not yet
been attained. Somewhere between these two limiting temperatures is a region
of isomerization processes with a character intermediate, in some sense,
between solid and liquid.

\begin{figure}
    \epsfxsize=4in
    \epsfysize=3in
    \centerline{\epsffile{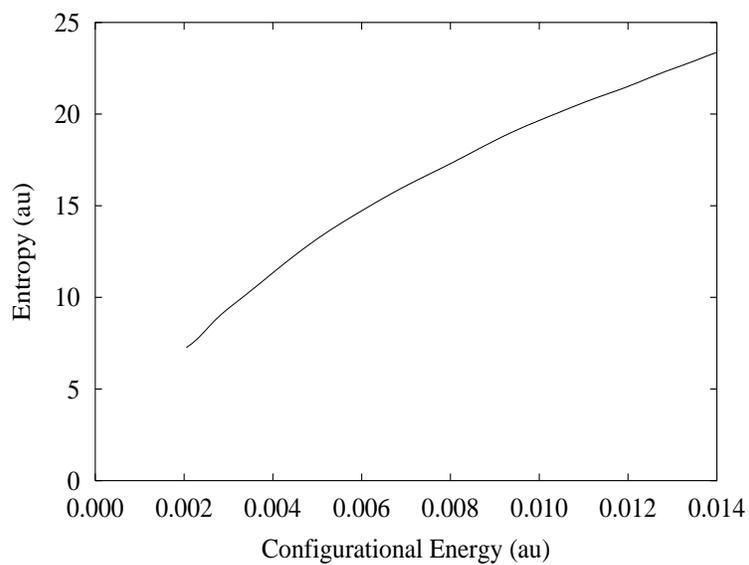}}   
    \caption {Ionic entropy of  Na$_8$   for  the KS-AI model extracted by
    the multiple histogram method.}
    \label {ks:na8:ent}
\end{figure}

\begin{figure}
    \epsfxsize=4in
    \epsfysize=3in
    \centerline{\epsffile{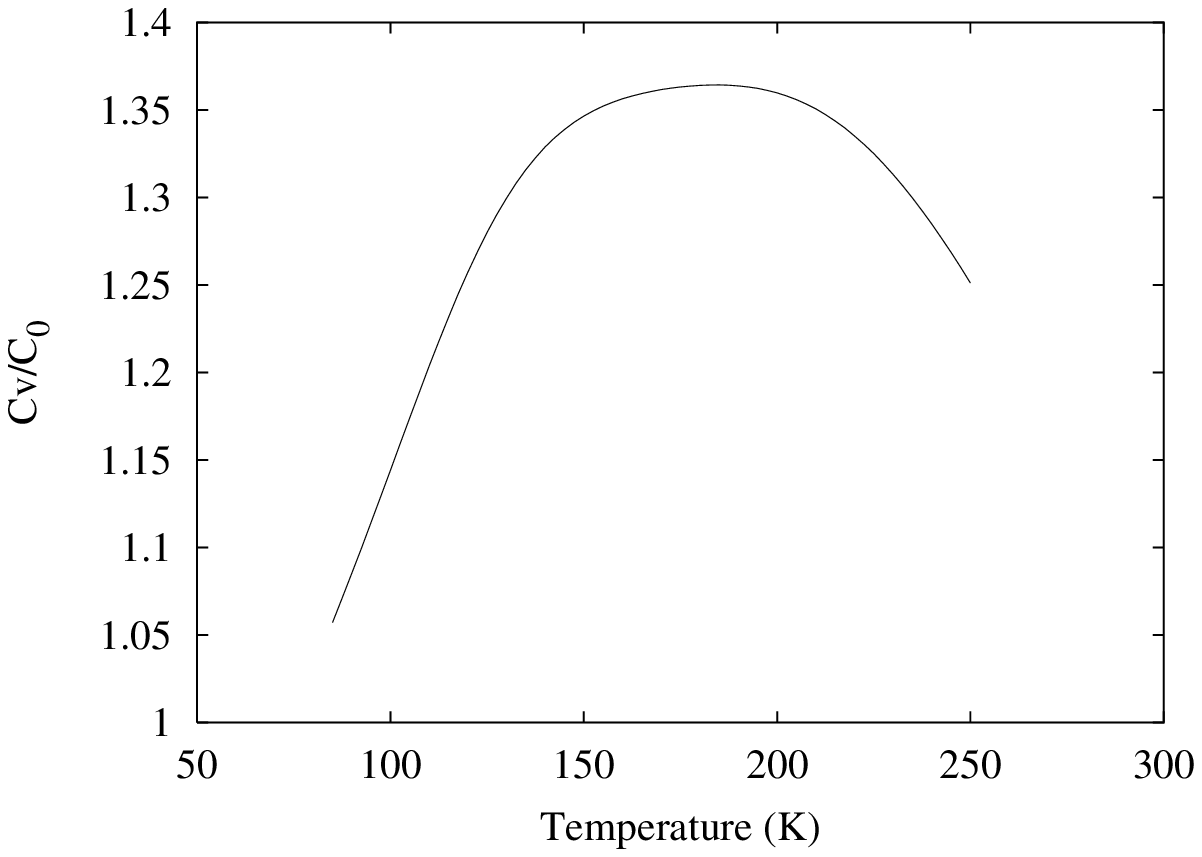}}   
    \caption {Canonical specific heat of Na$_8$ simulated using KS-AI.}
    \label {ks:na8:spht}
\end{figure}

\begin{figure}
    \epsfxsize=4in
    \epsfysize=3in
    \centerline{\epsffile{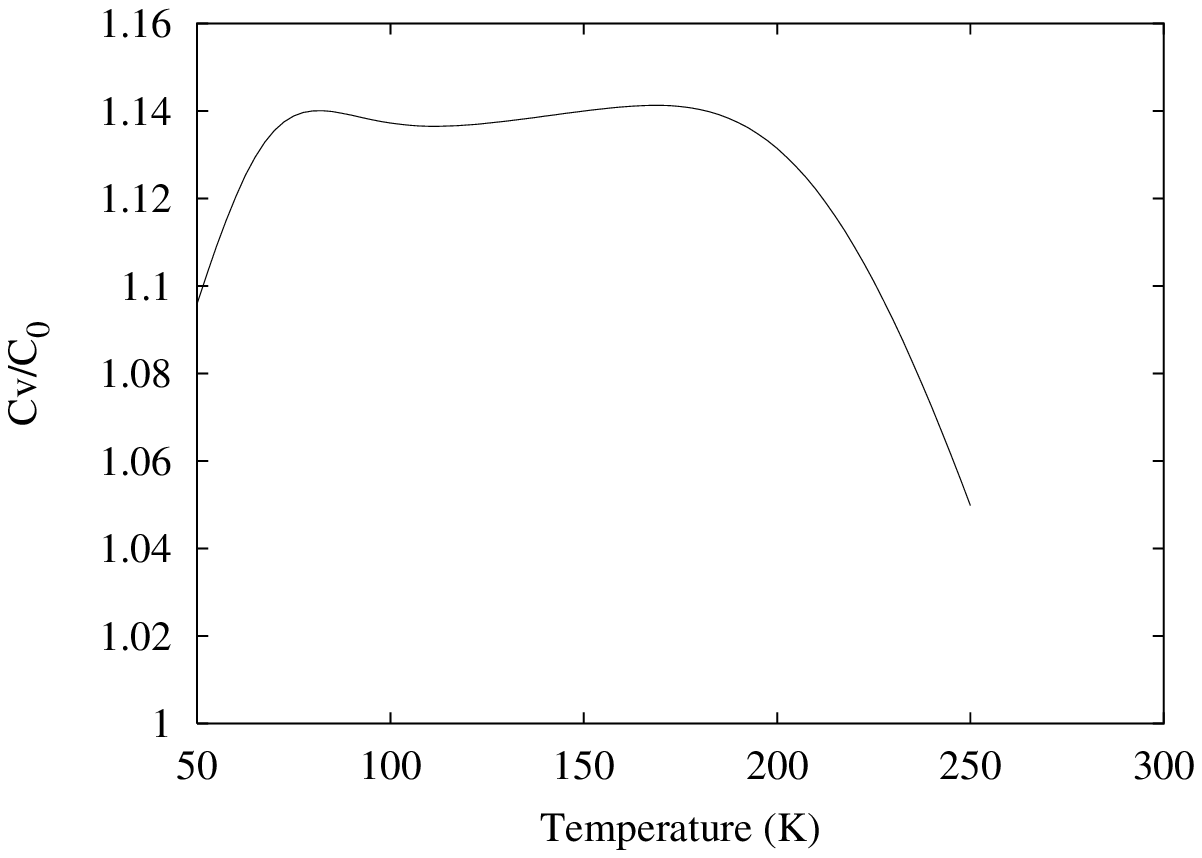}}   
    \caption {Canonical specific heat of  Na$_8$ simulated using DB-AI.}
    \label {dbmd:na8:spht}
\end{figure}

\begin{figure}
    \epsfxsize=4in
    \epsfysize=3in
    \centerline{\epsffile{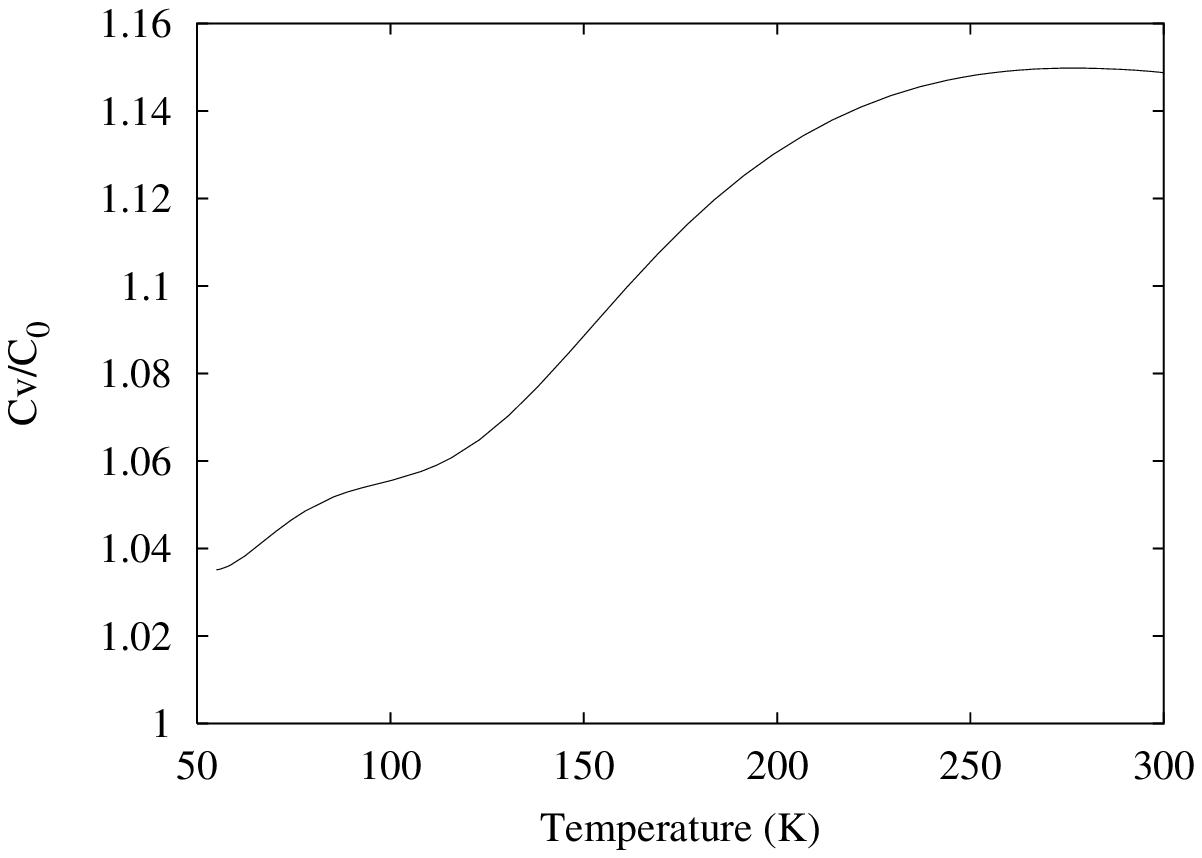}}
    \caption {Canonical specific heat of    Na$_8$ simulated using KS-SP.}
    \label {ks:na8:soft:spht}
\end{figure}

\begin{figure}
    \epsfxsize=4in
    \epsfysize=3in
    \centerline{\epsffile{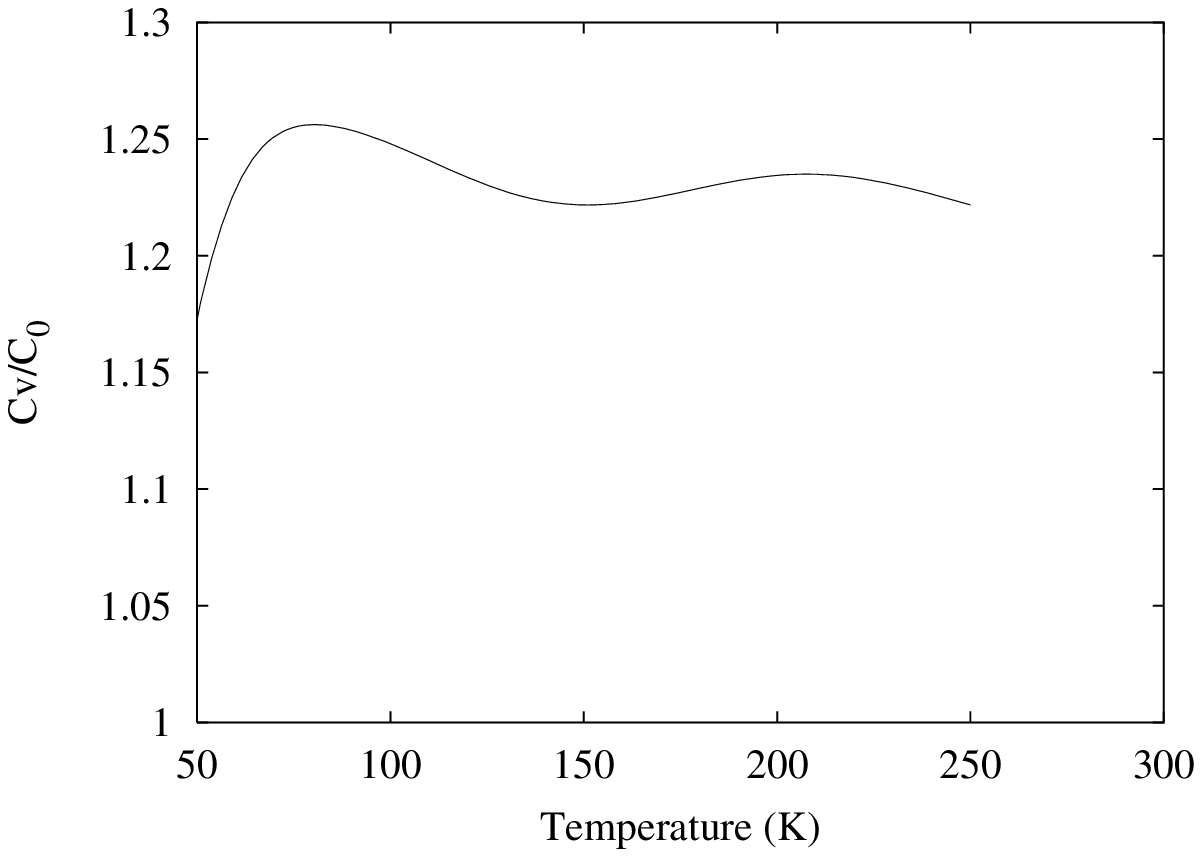}}
    \caption {Canonical specific heat of Na$_8$ simulated using DB-SP.}
    \label {dbmd:na8:soft:spht}
\end{figure}

\begin{figure}
    \epsfxsize=4in
    \epsfysize=3in
    \centerline{\epsffile{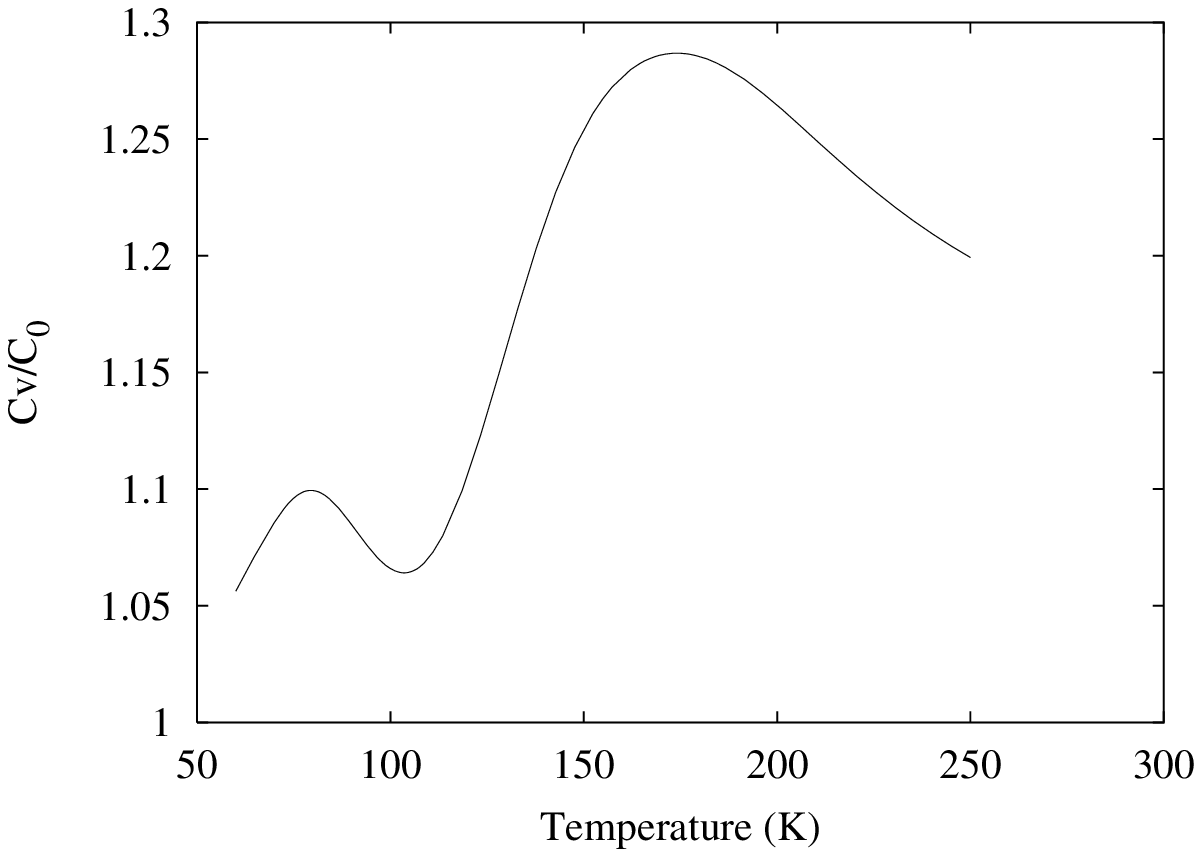}}
    \caption {Canonical specific heat of Na$_{20}$ simulated using KS-SP.}
    \label {ks:na20:soft:spht}
\end{figure}

\begin{figure}
    \epsfxsize=4in
    \epsfysize=3in
    \centerline{\epsffile{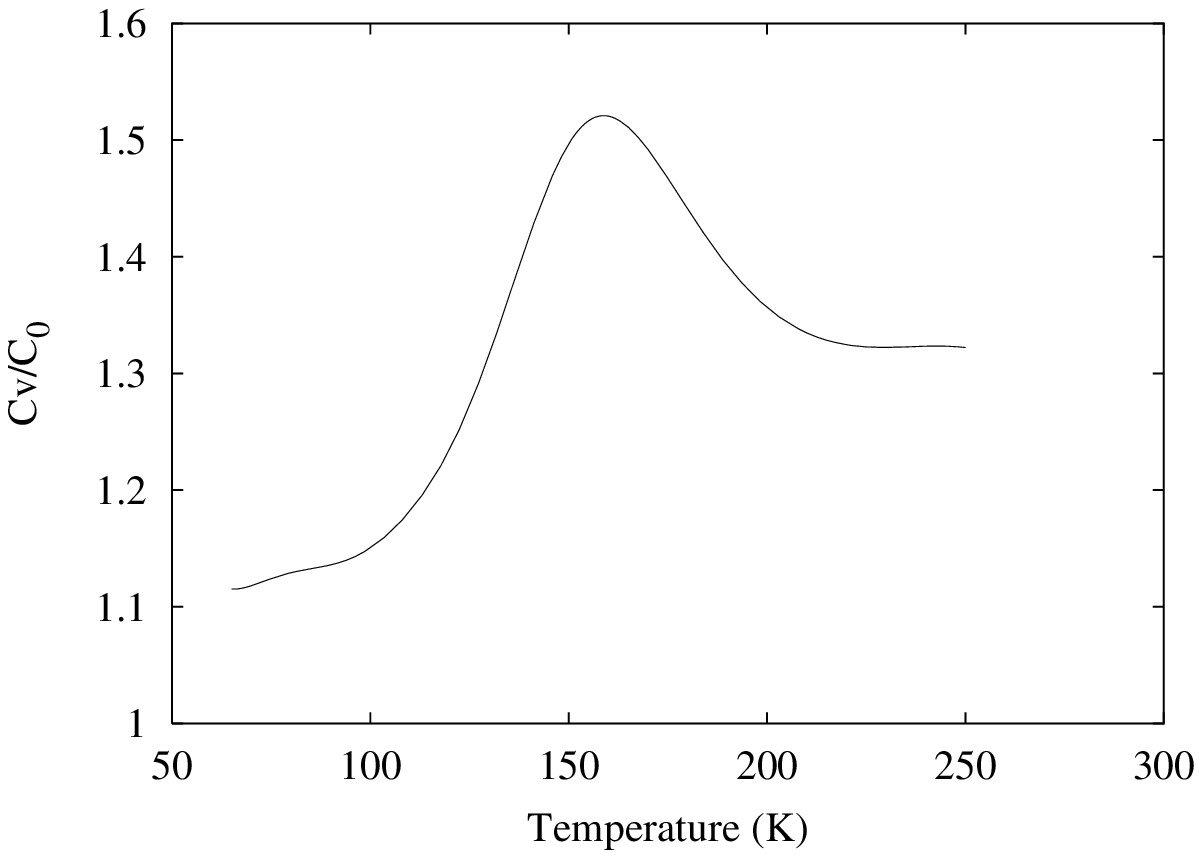}}
    \caption {Canonical specific heat of Na$_{20}$ simulated using DB-SP.}
    \label {dbmd:na20:soft:spht}
\end{figure}

\begin{figure}
    \epsfxsize=4in
    \epsfysize=3in
    \centerline{\epsffile{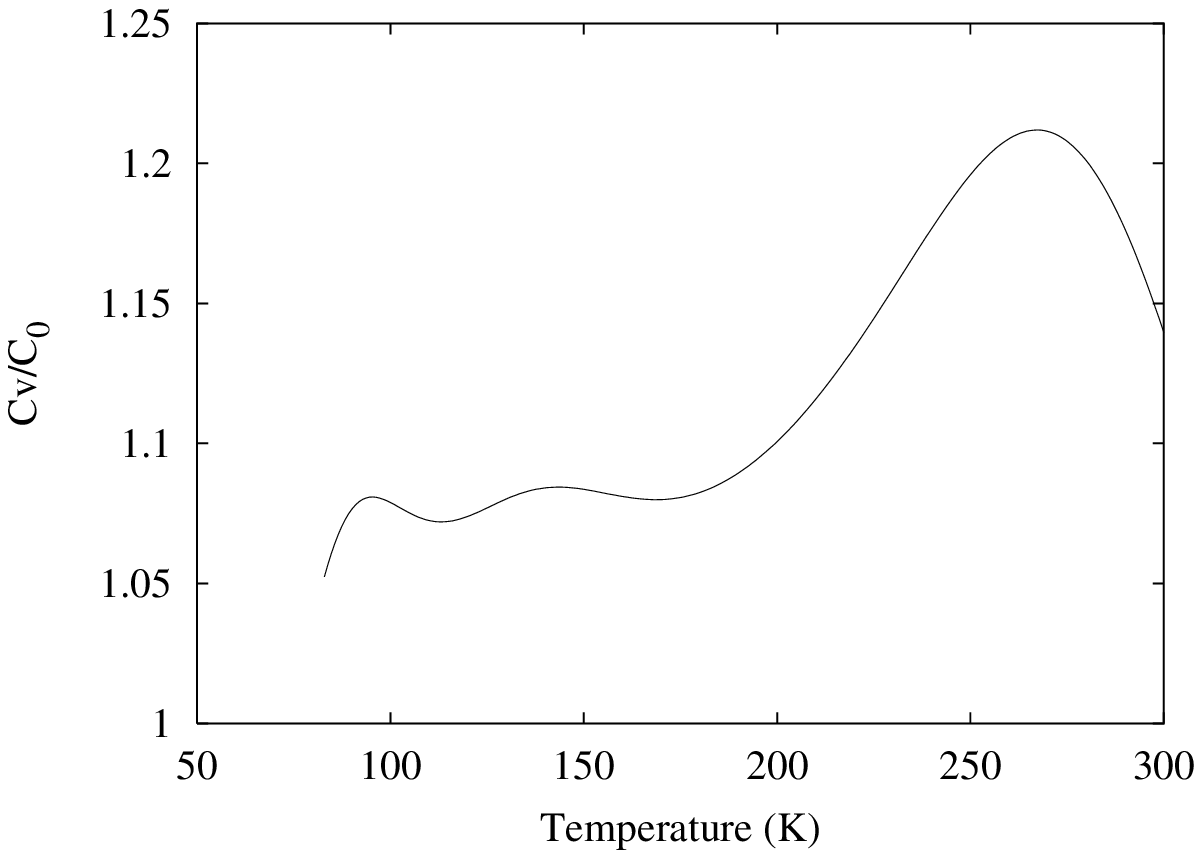}}   
    \caption {Canonical specific heat of Na$_{20}$ simulated using DB-AI.}
    \label {dbmd:na20:spht}
\end{figure}

The MH analysis may be used to probe further the thermodynamics of the
cluster in any particular model. One here extracts the ionic entropy $%
S(E)=\ln \Omega (E)$, which is a functional of the potential-energy surface (%
\ref{eqn:epot}). As expected, all entropy curves show a monotonic increase,
the curve for KS-AI, shown in Fig.\ \ref{ks:na8:ent}, being typical. The
canonical specific heats (\ref{eqn:specheat}) for Na$_{8}$ obtained via the
MH technique for all four models are shown in Figs.\ \ref{ks:na8:spht}--\ref
{dbmd:na8:soft:spht}. In general, all the Na$_{8}$ specific-heat curves show
broad peaks with widths over 100~K. The initial rise is around 70~K for both
DB models. In the KS models, the initial rise of the main peak for the SP
pseudopotential is at a higher temperature than for the AI\ pseudopotential,
namely, at 200~K compared to 150~K. However, the KS-SP model has a shoulder
feature around 80~K not visible in the KS-AI results. Turning to Na$_{20}$
in Figs.\ \ref{ks:na20:soft:spht}--\ref{dbmd:na20:spht}, we find main peaks
that are less broad than for Na$_{8}$, with a width generally somewhat less
than 100~K. In the DB models, the main peak for the AI pseudopotential is at
a higher temperature than for the SP pseudopotential, namely, at about 250~K
compared to about 150~K. If on the other hand we compare the KS-SP model
with the DB-SP\ model, we find that both main peaks occur at about the same
temperature. However, the KS-SP model has a ``premelting'' feature around
80~K that is more distinct than for DB-SP model.

It is difficult to draw simple, general conclusions from these observations
concerning the effect of the KS approach versus the DB approach, or the
effect of AI versus SP pseudopotentials. For example, for Na$_{8}$ in the
KS\ model, the SP pseudopotential gives a main peak at higher a temperature
than for the AI pseudopotential (if one ignores the small premelting feature
in the former), while for Na$_{20}$ in the DB model, it is the AI
pseudopotential that gives a peak at the higher temperature. For these small
cluster sizes, the precise form of the specific-heat curves can evidently be
very sensitive to the model used. One observes a similarly large variation
in specific-heat curves between the SMA and TB models reported in Ref.\ \cite
{calvo2}. 
Evidently, the important features of the potential-energy 
landscape can be rather sensitive to the model employed.

Some insight into the model-dependence of the potential-energy surface
may be gained from the energetic ordering of a selection of
possible isomers of Na$_{8}$.  We consider the dodecahedron $D_{2d}$ 
(the ground state in all DB and KS models of this work, as well as in the
SMA model \cite{calvo}), the capped pentagonal bipyramid 
$C_{s}$ (the ground state for LJ$_{8}$), and the stellated tetrahedron 
$T_{d}$. In the DB models and the SMA model, the $C_{s}$ structure forms 
a relatively low-lying excited isomer at 0.03--0.05~eV above the $D_{2d}$
ground state, while in our KS models and in the KS approach of 
Ref.\ \cite {andreoni}, the $C_{s}$ structure is unstable and
collapses to $D_{2d}$ upon relaxation.  On the other hand, the $T_{d}$ 
structure forms a higher-lying isomer at around 0.09--0.12~eV in the 
present DB and KS models, in the KS approach of Ref.\ \cite {andreoni}, 
and in the SMA model, while in the TB model \cite{calvo2} and in
an all-electron configuration-interaction approach,\cite{vlasta} the $T_{d}$
structure is  the ground state.  This illustrates how even for Na$_{8}$ the
ordering of isomers given by {\em ab initio} calculations is uncertain.  We
note that, while the heights of the barriers separating isomers are a
more important determining factor than the simple energy differences, the
existence of the low-lying $C_{s}$ isomer in the DB models, but not
in the KS models, is consistent with the lower-temperature shoulder of
the Na$_{8}$ specific heat curve in the DB models.

Our specific-heat curves are in general qualitatively quite similar to those
for the SMA potential:\cite{calvo,calvo2}  Na$_{8}$ in the SMA model has a
broad peak, and Na$_{20}$ a rather narrower peak with a small premelting feature
on the low-temperature side. On the other hand, there are some differences
with the TB specific-heat curves.\cite{calvo2}  For instance, the specific
heat of Na$_{8}$ has a somewhat narrower peak in the TB model than in the
DB, KS, or SMA models. However, as noted above, the ground-state
structure of Na$_{8}$ has a $T_{d}$ symmetry in the TB model, but a $D_{2d}$
symmetry in the DB, KS, and SMA models. Finally, Aguado {\em et. al}., \cite
{1:aguado} using a model quite similar to the present DB-AI model, give
{\em microcanonical} specific heats for Na$_{8}$ and Na$_{20}$, derived from
a trajectory-based analysis, that appear to disagree qualitatively with the
present DB and KS results (and with the SMA\ results.\cite{calvo,calvo2})
Their curve for Na$_{8}$ has a single narrow peak with a width less than 30~K
located around 110~K, while their curve for Na$_{20}$ has two distinct
peaks, each with a width less than 30~K and of similar height, located at 
about 110~K and 170~K. The precise reason for
the differences between their results and the present results is unclear at
present and requires further investigation. A re-evaluation of our own
results in the microcanonical ensemble shows that the change of ensemble is
insufficient to explain these differences, and we note that our DB
ground-state geometries agree with theirs. Given the similarity between
their DB model and ours, the discrepancies may be due simply to
methodological differences: in the present work we have derived the
specific-heat curves from a MH analysis, and have used longer sampling runs,
checking that the specific-heat curves are reasonably stable against the
addition of further data.
\section{Conclusions}

\label{conclusions}We have investigated the thermodynamics and melting of
the small clusters Na$_{8}$ and Na$_{20}$ using interionic potentials
derived from several DFT models for the valence electrons. The data have
been analyzed using a MH analysis, which is an efficient and reliable way of
probing the melting transition. Of the various DFT models, the most accurate
one considered here should be the KS-AI model. The other models involve
substituting less accurate electron kinetic-energy functionals $T[\rho ]$
(the DB approaches), or else soft, phenomenological pseudopotentials (the SP
approaches) in place of {\em ab initio} pseudopotentials, in each case with
a view to accelerating the calculation and permitting better statistics.
While there are qualitative similarities between the curves obtained from
the various models, we also observe substantial shifts in temperatures of
the main peaks and other features of the curves from model to model.

Concerning the choice of pseudopotential, while the SP pseudopotential is
known to predict ground-state geometries and certain other properties well,
as mentioned previously, it does not necessarily follow that energetic
barriers and other important features of the potential-energy surface
are well described. Given the substantial differences in the specific-heat
curves obtained from the SP and AI pseudopotentials, it would therefore seem
wise to prefer AI pseudopotentials. Note, however, that the soft
pseudopotential used here is highly phenomenological:
it lacks entirely a repulsive core, and deviates from the asymptotic value $%
-1/r$ for r$<\sim 3.7\,a_{0}$, which is well outside the physical core of
the Na$^{+}$ ion, r$<\sim 2.0\,a_{0}$. It may be possible to construct a
better soft pseudopotential that minimizes the difference with the
specific-heat curve obtained from an AI\ pseudopotential and yet still
yields a significantly cheaper calculation, thus permitting a very useful
gain in statistics in thermodynamic simulations. We are currently
investigating such possibilities.

As to the DB approach versus the full KS approach, we note that one
important approximation in the DB approach in its present form is its
difficulty in accounting for quantum shell effects accurately. The two DB
forms for $T[\rho ]$ considered here yield energies that vary smoothly with
cluster size $N$ according to a liquid-drop formula (as shown explicitly in
Ref.\ \cite{steve} for the extended Thomas-Fermi functional), without
showing the fluctuations in energy associated with quantum shell closures of
the valence electrons. Further, while ground-state geometries and other
properties of closed-shell systems can be predicted rather successfully by
the DB approach \cite{steve,dbmd}, it has trouble reproducing Jahn-Teller
distortions in open-shell systems. Now, according to the Hohenberg-Kohn
theorems, it should in principle be possible to find a DB functional $T[\rho
]$ that fully incorporates such quantum shell effects in finite systems. It
appears that if the DB methods are to yield reliable, quantitative
information, then better electronic kinetic-energy functionals along these
lines are required. Fortunately, significant progress is being made in this
direction, and a number of researchers have already proposed DB
kinetic-energy functionals incorporating electronic shell effects
approximately.

There is a special reason for paying careful attention to electronic shell
effects in studies of melting that was mentioned earlier, related to the
experiments of Haberland {\em et al.} \cite{schmidt}. The fact that
prominent maxima in the melting point occur for sizes that are intermediate
between geometric shell closures and electronic shell closures suggests that
there is an important interplay between geometric effects and quantum shell
effects. To understand this phenomenon more closely, we are currently
considering Monte-Carlo sampling methods combined with various approximate
KS schemes, with a view to extending the KS calculations presented in this
work to larger sizes, within the range of the Haberland {\em et al.} data.
\section*{Acknowledgements}

   We gratefully acknowledge the support of the Indo-French Center for
   the  Promotion of Advanced Research (New
   Delhi)   / Centre Franco-Indien  pour la  Promotion de la Recherche
   Avanc\'ee under contract 1901-1.  One of us (AMV) 
   acknowledges the hospitality of the
   CEA, Grenoble, France.  AMV is grateful to CSIR, New Delhi, India for
   their research fellowship.
   Thanks  are also due   to Matt Freigo  and Stephen  Johnson  for an
   excellent FFT library.

\section*{Figures}

%
%

%
%

\begin{thebibliography}{99}
\bibitem {units}
Atomic  units are  used in the paper,  unless  otherwise explicitly 
stated.
\end{thebibliography}
\end{document}